\documentclass[12pt]{iopart}
\usepackage[utf8]{inputenc}
\usepackage{graphicx}
\usepackage{cite}
\usepackage[none]{hyphenat}
\usepackage{amssymb}
\usepackage{float}
\usepackage{lineno}

\usepackage[bigfiles]{pdfbase}
\ExplSyntaxOn
\NewDocumentCommand\embedvideo{smm}{
  \group_begin:
  \leavevmode
  \tl_if_exist:cTF{file_\file_mdfive_hash:n{#3}}{
    \tl_set_eq:Nc\video{file_\file_mdfive_hash:n{#3}}
  }{
    \IfFileExists{#3}{}{\GenericError{}{File~`#3'~not~found}{}{}}
    \pbs_pdfobj:nnn{}{fstream}{{}{#3}}
    \pbs_pdfobj:nnn{}{dict}{
      /Type/Filespec/F~(#3)/UF~(#3)
      /EF~<</F~\pbs_pdflastobj:>>
    }
    \tl_set:Nx\video{\pbs_pdflastobj:}
    \tl_gset_eq:cN{file_\file_mdfive_hash:n{#3}}\video
  }
  \pbs_pdfobj:nnn{}{dict}{
    /Type/RichMediaInstance/Subtype/Video
    /Asset~\video
    /Params~<</FlashVars (
      source=#3&
      skin=SkinOverAllNoFullNoCaption.swf&
      skinAutoHide=true&
      skinBackgroundColor=0x5F5F5F&
      skinBackgroundAlpha=0
    )>>
  }
  \pbs_pdfobj:nnn{}{dict}{
    /Type/RichMediaConfiguration/Subtype/Video
    /Instances~[\pbs_pdflastobj:]
  }
  \pbs_pdfobj:nnn{}{dict}{
    /Type/RichMediaContent
    /Assets~<<
      /Names~[(#3)~\video]
    >>
    /Configurations~[\pbs_pdflastobj:]
  }
  \tl_set:Nx\rmcontent{\pbs_pdflastobj:}
  \pbs_pdfobj:nnn{}{dict}{
    /Activation~<<
      /Condition/\IfBooleanTF{#1}{PV}{XA}
      /Presentation~<</Style/Embedded>>
    >>
    /Deactivation~<</Condition/PI>>
  }
  \hbox_set:Nn\l_tmpa_box{#2}
  \tl_set:Nx\l_box_wd_tl{\dim_use:N\box_wd:N\l_tmpa_box}
  \tl_set:Nx\l_box_ht_tl{\dim_use:N\box_ht:N\l_tmpa_box}
  \tl_set:Nx\l_box_dp_tl{\dim_use:N\box_dp:N\l_tmpa_box}
  \pbs_pdfxform:nnnnn{1}{1}{}{}{\l_tmpa_box}
  \pbs_pdfannot:nnnn{\l_box_wd_tl}{\l_box_ht_tl}{\l_box_dp_tl}{
    /Subtype/RichMedia
    /BS~<</W~0/S/S>>
    /Contents~(embedded~video~file:#3)
    /NM~(rma:#3)
    /AP~<</N~\pbs_pdflastxform:>>
    /RichMediaSettings~\pbs_pdflastobj:
    /RichMediaContent~\rmcontent
  }
  \phantom{#2}
  \group_end:
}
\ExplSyntaxOff

\begin{document}

\title[Double-layer and reverse discharge in BP-HiPIMS]{On double-layer and reverse discharge creation during long positive voltage pulses in a bipolar HiPIMS discharge}

\author{A D Pajdarová$^1$, T Kozák$^1$, T Tölg$^1$, and J Čapek$^1$}
\address{$^1$ Department of Physics and NTIS -- European Centre of Excellence, University of West Bohemia, Univerzitní 8, 301 00 Plzeň, Czech Republic}
\ead{adp@kfy.zcu.cz}

\begin{abstract}
Time-resolved Langmuir probe diagnostics at the discharge centerline and at three distances from the target ($35\,\mathrm{mm}$, $60\,\mathrm{mm}$, and $100\,\mathrm{mm}$) was carried out during long positive voltage pulses (a duration of $500\,\mathrm{\mu s}$ and a preset positive voltage of $100\,\mathrm{V}$) in bipolar High-Power Impulse Magnetron Sputtering of a Ti target (a diameter of $100\,\mathrm{mm}$) using an unbalanced magnetron. Fast-camera spectroscopy imaging recorded light emission from Ar and Ti atoms and singly charged ions during positive voltage pulses. It was found that during the long positive voltage pulse, the floating and the plasma potentials suddenly decrease, which is accompanied by the presence of anode light located on the discharge centerline between the target center and the magnetic null of the magnetron's magnetic field. These light patterns are related to the ignition of a reverse discharge, which leads to the subsequent rise in the plasma and the floating potentials. The reversed discharge is burning up to the end of the positive voltage pulse, but the plasma and floating potentials have lower values than the values from the initial part of the positive voltage pulse. Secondary electron emission induced by the impinging Ar$^+$ ions to the grounded surfaces in the vicinity of the discharge plasma together with the mirror configuration of the magnetron magnetic field are identified as the probable causes of the charge double-layer structure formation in front of the target and the ignition of the reverse discharge.
\end{abstract}


\submitto{\PSST}

\maketitle

\section{Introduction}

High-power Impulse Magnetron Sputtering (HiPIMS) discharge is a development of the conventional DC magnetron sputtering technique when high power is delivered to the magnetron target in the form of periodically repeating voltage and current pulses with a relatively low duty cycle (typically $\lesssim 20\,\%$) \cite{Sarakinos2010, Gudmundsson2012, Anders2017}. This is connected with several significant benefits for film deposition: (1) strong sputtering of atoms from the magnetron target, (2) huge and highly ionized fluxes of atoms to the substrate, (3) augmented energies of ions flowing onto the substrate, (4) enhanced dissociation of reactive gas molecules, and (5) reduced poisoning (coverage with the reactive gas compounds) of the target surface during the reactive deposition \cite{Sarakinos2010, Gudmundsson2012, Vlcek2013, Anders2017, Capek2017}. Unfortunately, the deposition rate of HiPIMS discharges is usually lower or comparable with the conventional DC magnetron sputtering at similar average power densities, mainly resulting from the return of ions to the target \cite{Gudmundsson2012, Anders2017}. So, it is advantageous for the film deposition to release these trapped ions from the target vicinity toward the substrate \cite{Capek2013}.

One possibility how to push ions in the direction toward the substrate after the end of a negative voltage pulse (NP), where the target sputtering takes place, is to apply a positive voltage pulse (PP) to the target after the NP termination \cite{Nakano2010a}. This technique is usually called bipolar HiPIMS (BP-HiPIMS). It was used successfully to improve the deposition of different types of films, e.g., to increase the deposition rate \cite{Wu2018, Ganesan2023}, to improve hardness \cite{Eichenhofer2017, Viloan2019, Tiron2019, Batkova2020b, Fernandez-Martinez2022}, strengthen film adhesion \cite{Wu2018, Velicu2019, Cho2023}, for film densification \cite{Fernandez-Martinez2022, Quillin2022}, to smoothen the film surface \cite{Tiron2019, Velicu2019}, to increase grain sizes \cite{Batkova2020b, Quillin2022}, to rise the content of sp$^3$ bonds in diamond-like carbon films \cite{Tiron2019, Santiago2019, Gaines2019, Ganesan2023}, and to minimize porosity \cite{Quillin2022}.

The effect of PP on the plasma behavior has been analyzed using many diagnostic techniques in recent years. The results of time-averaged energy-resolved mass spectroscopy \cite{Hippler2019, Keraudy2019, Santiago2019, Velicu2019, Viloan2019, Hippler2020, Hippler2020a, Kozak2020, Avino2021}, and those of retarding field energy analyzers \cite{Han2022, Walk2022a} are quite similar among different sputtering systems. They show that the ion energy distribution function (IEDF) is enhanced by a group of ions whose energies are around a value corresponding to the applied voltage during PP. Time- and energy-resolved mass spectroscopy confirmed that these ions originate from the PP \cite{Kozak2020, Hippler2020a}.

Laser-induced fluorescence applied during BP-HiPIMS sputtering of a Ti target \cite{Britun2018} showed that the ground state density of Ti$^+$ ions is significantly reduced during PP in comparison with their density without PP as a result of Ti$^+$ ions signiﬁcant acceleration in the target-to-substrate direction. Time-resolved optical emission spectroscopy (OES) was also carried out in BP-HiPIMS discharges \cite{Hippler2019, Hippler2020, Klein2023} with the result that only argon atom emission lines resume during PP if a sufficiently high positive voltage ($U_+ \gtrsim 25\,\mathrm{V}$) is applied.

The results of Langmuir or emissive probe measurements from different sputtering systems do not agree. Generally speaking, the results may be divided into two scenarios \cite{Han2022}, which are also connected with explaining how and where the ions gain high energies. The first one is that a charge double-layer (DL) potential structure is formed in front of the target biased to positive voltages \cite{Keraudy2019, Velicu2019, Tiron2020}. The ions then gain energy when they cross the DL potential structure from the DL side closer to the target, where the plasma potential is higher (H-side), to the DL side closer to the substrate with substantially lower plasma potential (L-side). Thus, it is believed that the DL structure makes it possible to enhance the ion bombardment even on insulating substrates. Moreover, the electrons flowing from the L-side to the H-side gain enough energy to ionize atoms on the H-side of the DL structure. The DL structure usually forms gradually in time \cite{Velicu2019, Tiron2020}, and a distinct decrease in the plasma potential between the H-side and the L-side may be registered somewhere at a distance between the target and the substrate.

The second scenario is that the ions gain energy mainly in the potential drop across the substrate sheath. In this case, the plasma potential rises very quickly (on time scales substantially lower or at most on the order of $\mathrm{\mu s}$) to positive values in the plasma volume after applying the positive voltage to the target. This leads to the formation of a sheath close to the substrate (grounded or negatively biased) where the ions are accelerated onto the growing film \cite{Hippler2020, Pajdarova2020, Law2021}.

The experimental data well support both scenarios. Identifying which discharge parameters govern the selection of the above-mentioned scenarios is still undergoing investigation. It is believed that one possible candidate that controls plasma behavior during PP is the magnetic field topology of the magnetron used (a balanced vs. unbalanced magnetic field configuration) \cite{Pajdarova2020}.

In recent years, Particle-in-Cell (PIC) simulations have been carried out to shed light on plasma processes taking place during PP \cite{Avino2021, Han2022}. The sputtering plasma must be simplified considerably to reduce the actual discharge's complexity and to achieve reasonable computation times. Only electrons, argon atoms, and argon ions are traced during these PIC simulations, and a reduced set of reactions among these particles is considered. Despite these substantial simplifications, the PIC simulations provide valuable information on plasma properties very close to the target, e.g., the plasma potential and the resulting electric field, the local breach of charge quasineutrality, and the mapping of particle fluxes, where diagnostics methods cannot be easily applied.

Kozák {\it et al.} \cite{Kozak2020} showed that for longer PP (a duration of $200\,\mathrm{\mu s}$), a third broad peak in IEDF appears, with energy reduced by roughly $15\,\mathrm{eV}$ compared to the highest peak. This third peak corresponds in time with the drop of the plasma potential (measured by a Langmuir probe) and a sudden increase of plasma emission from Ar atoms (detected by fast-camera imaging) in a form resembling a "light bulb" located at the discharge centerline near the target. This "reignition" of discharge was named "reverse discharge" (RD). Similar drops in the plasma and/or floating potentials were also detected by other authors in different BP-HiPIMS systems \cite{Hippler2020, Avino2021, Law2021, Zanaska2022}. Usually, the drop in potentials is, after some time, followed by their increase to new, almost stationary values, which are lower than those measured after the PP initiation. This event was named drop and rise (D\&R) \cite{Avino2021}. Similar light-emission structures at the discharge centerline near the target were also confirmed by other authors in the discharge whole light \cite{Law2023} or in the emission of Ar atoms \cite{Klein2023}. This shows that the presence of RD in BP-HiPIMS may be common. Since RD may significantly influence the energy and composition of ion fluxes onto the growing film, this phenomenon is worth a detailed study.

\section{Experimental details and data processing}

\begin{figure}
    \centering
    \includegraphics{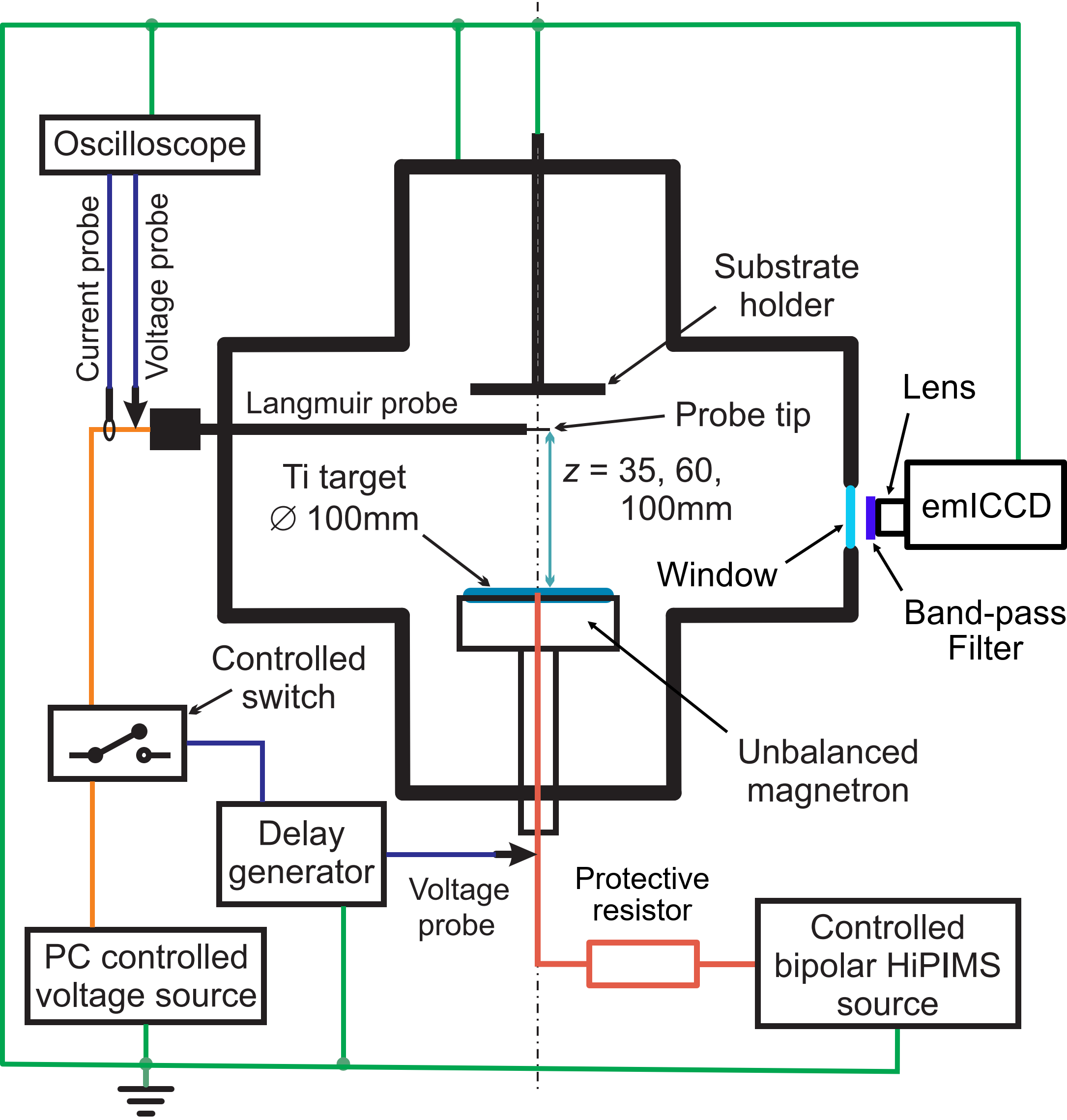}
    \caption{A schematic diagram of the experimental system used.}
    \label{f:system}
\end{figure}

Our experimental system for diagnosing BP-HiPIMS is depicted schematically in \fref{f:system}. A more detailed description of this system and the processing of Langmuir probe data can be found in Ref.~\cite{Pajdarova2020}. Here, we will mention information that is needed for understanding of presented results.

\begin{figure}
    \centering
    \includegraphics{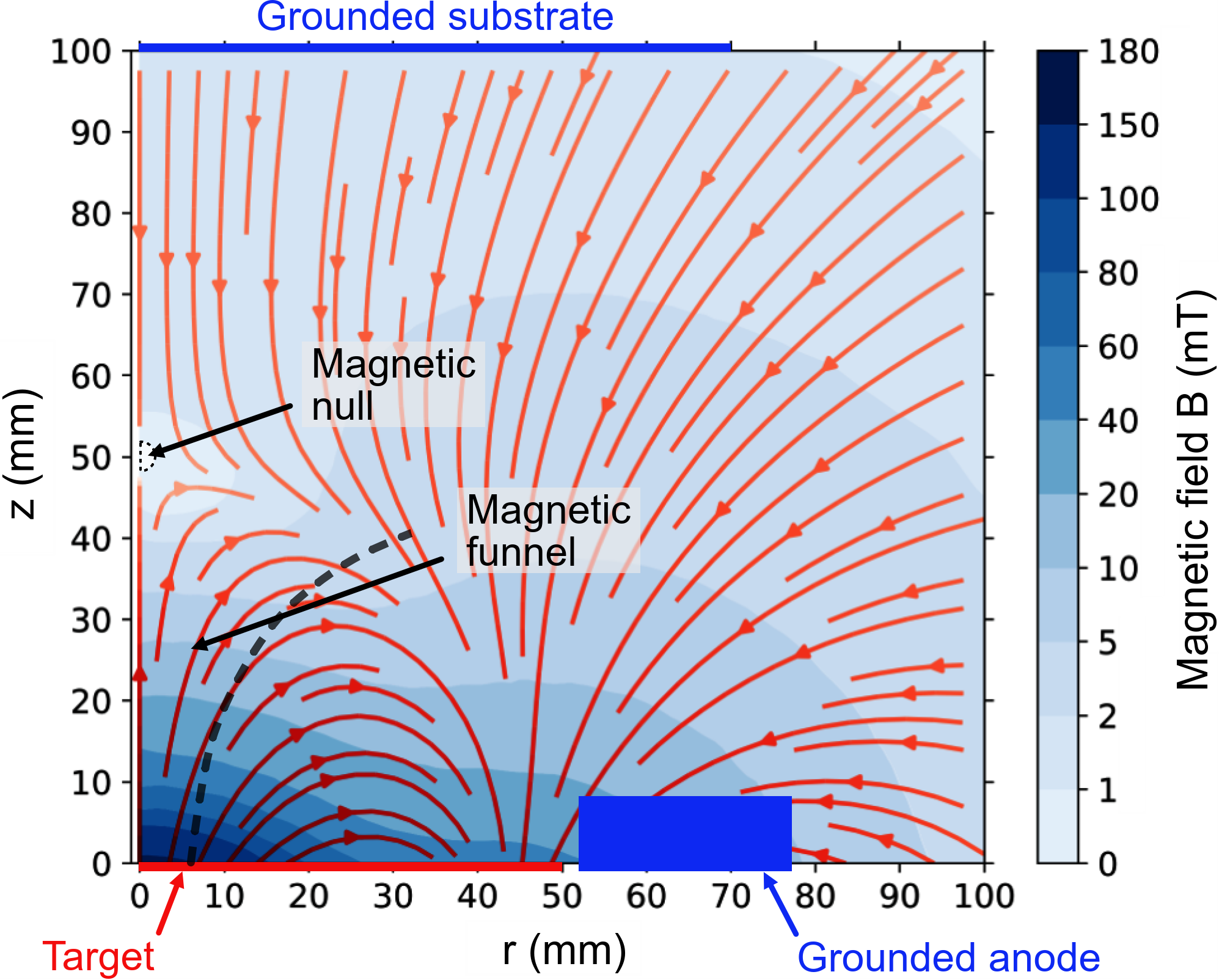}
    \caption{The magnetic field of the magnetron, where $r$ is the horizontal distance from the discharge centerline and $z$ is the vertical distance from the target surface. The positions of the target, protruding grounded anode, grounded substrate, magnetic funnel, and magnetic null are also depicted.}
    \label{f:magfield}
\end{figure}

The vacuum vessel was made from a DN\,200 ISO-K 6-way cross piping, and it was evacuated by a turbomolecular pump backed up by a scroll pump down to $5 \times 10^{-5}\,\mathrm{Pa}$. A self-built BP-HiPIMS power unit operating in the constant voltage mode was used for the powering of a planar circular magnetron with a Ti target (a diameter of $100\,\mathrm{mm}$ and a thickness of $6\,\mathrm{mm}$). The magnetron voltage and discharge currents were monitored by voltage and current probes connected directly to the magnetron. Only one regime with a NP duration of $t_- = 100\,\mathrm{\mu s}$, a PP duration of $t_+ = 500\,\mathrm{\mu s}$, a delay between the NP end and the PP initiation of $t_\mathrm{D} = 20\,\mathrm{\mu s}$, a PP voltage amplitude of $U_+ = 100\,\mathrm{V}$, and a process gas (argon) pressure of $p = 1\,\mathrm{Pa}$ is examined here. The magnetic field of the magnetron (see \fref{f:magfield}) was configured as a type-2 unbalanced field with the magnetic null at the axis of symmetry at the distance of $50\,\mathrm{mm}$ from the target. The approximative position of the magnetic funnel, where electrons are not strongly confined near the target by the magnetic field during NP, is also depicted in the figure.

The tip of the Langmuir probe (a length of $10\,\mathrm{mm}$ and a diameter of $0.15\,\mathrm{mm}$) made of tungsten was parallel with the target surface. It was positioned at the discharge centerline at the distances $z = 35\,\mathrm{mm}$, $60\,\mathrm{mm}$, and $100\,\mathrm{mm}$ from the target (see \fref{f:system}). A PC-controlled voltage source biased the probe via a MOSFET switch, which allowed us to avoid the probe tip overheating during NP owing to high electron currents at the high probe bias voltages (up to $200\,\mathrm{V}$ to the ground). The switch was turned on $5\,\mathrm{\mu s}$ before the PP initiation and turned off $100\,\mathrm{\mu s}$ after the PP end. From the waveforms of the probe current and voltage (measured between the switch and the probe tip) recorded by an oscilloscope for $128$ periods of discharge pulses, the probe current-voltage (IV) characteristics were reconstructed every $0.1\,\mathrm{\mu s}$ during PP. These IV characteristics were averaged to achieve a time resolution of $0.5\,\mathrm{\mu s}$ and $1\,\mathrm{\mu s}$ during PP and RD, respectively. The effects of the magnetic field on the electron current collected by the probe tip may be neglected as the magnetic field lines are almost perpendicular to the surface of the probe tip, and the magnetic field is relatively weak even at the distance of $35\,\mathrm{mm}$ from the target.

\begin{figure}
    \centering
    \includegraphics{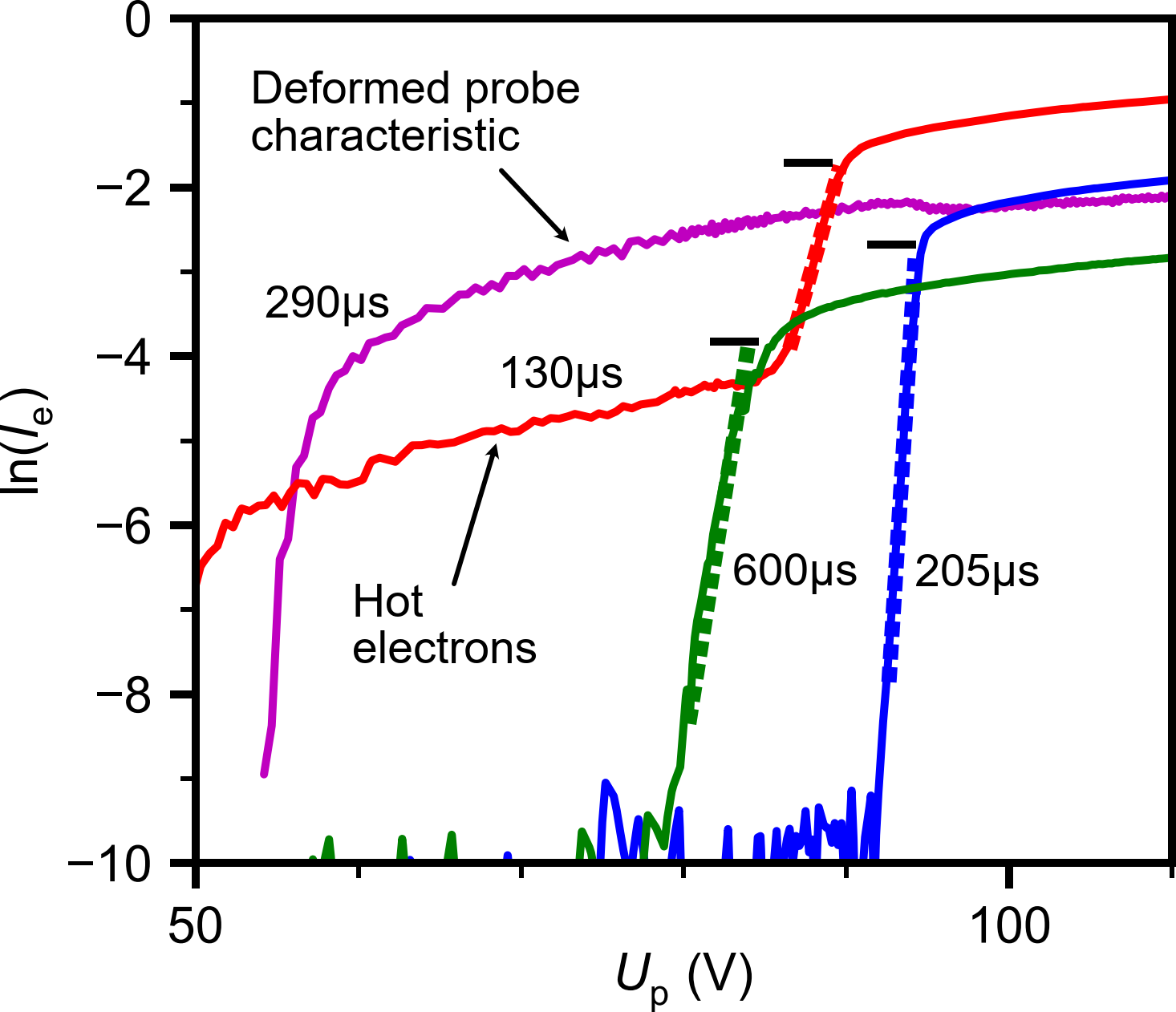}
    \caption{The natural logarithms of the electron probe currents, $I_\mathrm{e}$, (in amps) measured at the probe potentials, $U_\mathrm{p}$, for selected times during PP at the distance of $100\,\mathrm{mm}$ from the target. The bold broken lines show fits of equation \eref{e:Ie} in the logarithmic form to the measured data. The short horizontal lines mark the natural logarithm of the electron saturation currents determined by the fitting procedure.}
    \label{f:IVchar}
\end{figure}

\Fref{f:IVchar} shows the measured probe electron currents at the selected times of measurement $t_\mathrm{m} = 130\,\mathrm{\mu s}$ (after the PP initiation), $205\,\mathrm{\mu s}$ (during the first floating potential drop), $290\,\mathrm{\mu s}$ (during the RD ignition), and $600\,\mathrm{\mu s}$ (before the PP end) after the NP initiation ($t = 0\,\mathrm{\mu s}$) at the distance of $100\,\mathrm{mm}$ from the target. All the characteristics had the usual shape and were smooth, except several of them measured around the RD ignition (see the IV characteristic at $t_\mathrm{m} = 290\,\mathrm{\mu s}$ in \fref{f:IVchar}), which were deformed. Only the floating potential could be determined in these cases. Similarly deformed IV characteristics were also detected in Ref. \cite{Avino2021} around the RD ignition. The floating potential, $V_\mathrm{f}$, was determined as the probe potential $U_\mathrm{p}$ where the current of the probe was zero ($I_\mathrm{p}(V_\mathrm{f}) = 0\,\mathrm{A}$). As the IV characteristics are dominated by a group of almost Maxwellian electrons (except the first few $\mathrm{\mu s}$ at the very beginning of PP, where also hot electrons with approximately $10\times$ lower density are detected, see $t_\mathrm{m} = 130\,\mathrm{\mu s}$ in \fref{f:IVchar}), we determine the plasma potential, $V_\mathrm{p}$, the electron temperature, $T_\mathrm{e}$, and the electron density, $n_\mathrm{e}$, by fitting the measured probe current in the transient region ($U_\mathrm{p}$ in the upper two-thirds of an interval between $V_\mathrm{f}$ and $V_\mathrm{p}$) to the classical formula \cite{Chung1975}
\begin{equation}
I_\mathrm{e}(U_\mathrm{p}) = I^*_\mathrm{e} \exp \left[ -\frac{e (V_\mathrm{p} - U_\mathrm{p})}{k T_\mathrm{e}} \right],
\label{e:Ie}
\end{equation}
where $I^*_\mathrm{e} = 0.25 A e n_\mathrm{e} (8 k T_\mathrm{e} / \pi m_\mathrm{e})^{0.5}$ is the electron saturation current, $A$ is the probe tip area, $e$ is the elementary charge, $k$ is the Boltzmann constant, and $m_\mathrm{e}$ is the electron mass. For better results, the natural logarithm of equation \eref{e:Ie} was fitted to the natural logarithm of the measured probe current with a removed ion current component. The initial value of $V_\mathrm{p}$, around which the fitting procedure is performed, is determined as the $U_\mathrm{p}$ value where the $\mathrm{d} I\mathrm{^s_p} / \mathrm{d} U_\mathrm{p}$ is maximum. The quantity $I\mathrm{^s_p}$ is the smoothed probe current produced by the second-order Savitzky--Golay filter with an automatically determined number of points. This procedure of IV characteristics evaluation was automated by a Python script, which gave us consistent results. The average and maximum relative differences of fitted $I_\mathrm{e}$ to the measured values were lower than $2\,\%$ and $5\,\%$ during the initial part of PP, $3\,\%$ and $7\,\%$ before the RD ignition, and $7\,\%$ and $12\,\%$ during RD, respectively.

The light emission of plasma during PP was recorded by an emICCD camera (PI-MAX 4 with SR Intensifier, Princeton Instruments), which can amplify the light signal by a light intensifier and by multiplying electrons on the chip simultaneously. The camera was equipped with a UV lens (a focal length of $25\,\mathrm{mm}$) with an appropriate band-pass filter in front of it: $520\,\mathrm{nm}$ with the FWHM of $10\,\mathrm{nm}$ for Ti atoms (dominant transmitted emission lines at $517.37\,\mathrm{nm}$, $519.30\,\mathrm{nm}$, and $521.04\,\mathrm{nm}$), $334\,\mathrm{nm}$ with the FWHM of $10\,\mathrm{nm}$ for Ti$^+$ ions (dominant transmitted emission lines at $332.29\,\mathrm{nm}$, $334.19\,\mathrm{nm}$, $334.94\,\mathrm{nm}$, $336.12\,\mathrm{nm}$ and $337.28\,\mathrm{nm}$), $811\,\mathrm{nm}$ with the FWHM of $3\,\mathrm{nm}$ for Ar atoms (dominant transmitted emission lines at $810.37\mathrm{nm}$ and $811.53\,\mathrm{nm}$), and $488\,\mathrm{nm}$ with the FWHM of $3\,\mathrm{nm}$ for Ar$^+$ ions (dominant transmitted emission lines at $487.99\,\mathrm{nm}$ and $488.90\,\mathrm{nm}$). During PP, both amplification methods of the camera were used (a theoretical amplification of 10000) to capture images each $5\,\mathrm{\mu s}$ with a gate width of $5\,\mathrm{\mu s}$ to cover the whole PP. Moreover, for Ti atoms, Ti$^+$ ions, and Ar$^+$ ions, charge accumulation on the chip had to be used as the radiation of RD is very weak. For each measured atomic and ionic species, the light intensity during PP was normalized independently so that $1$ corresponds to the most intense light emission from a given species during PP. This allows us to emphasize the structure of light patterns in the discharge for all monitored species.

\begin{figure}
    \centering
    \includegraphics{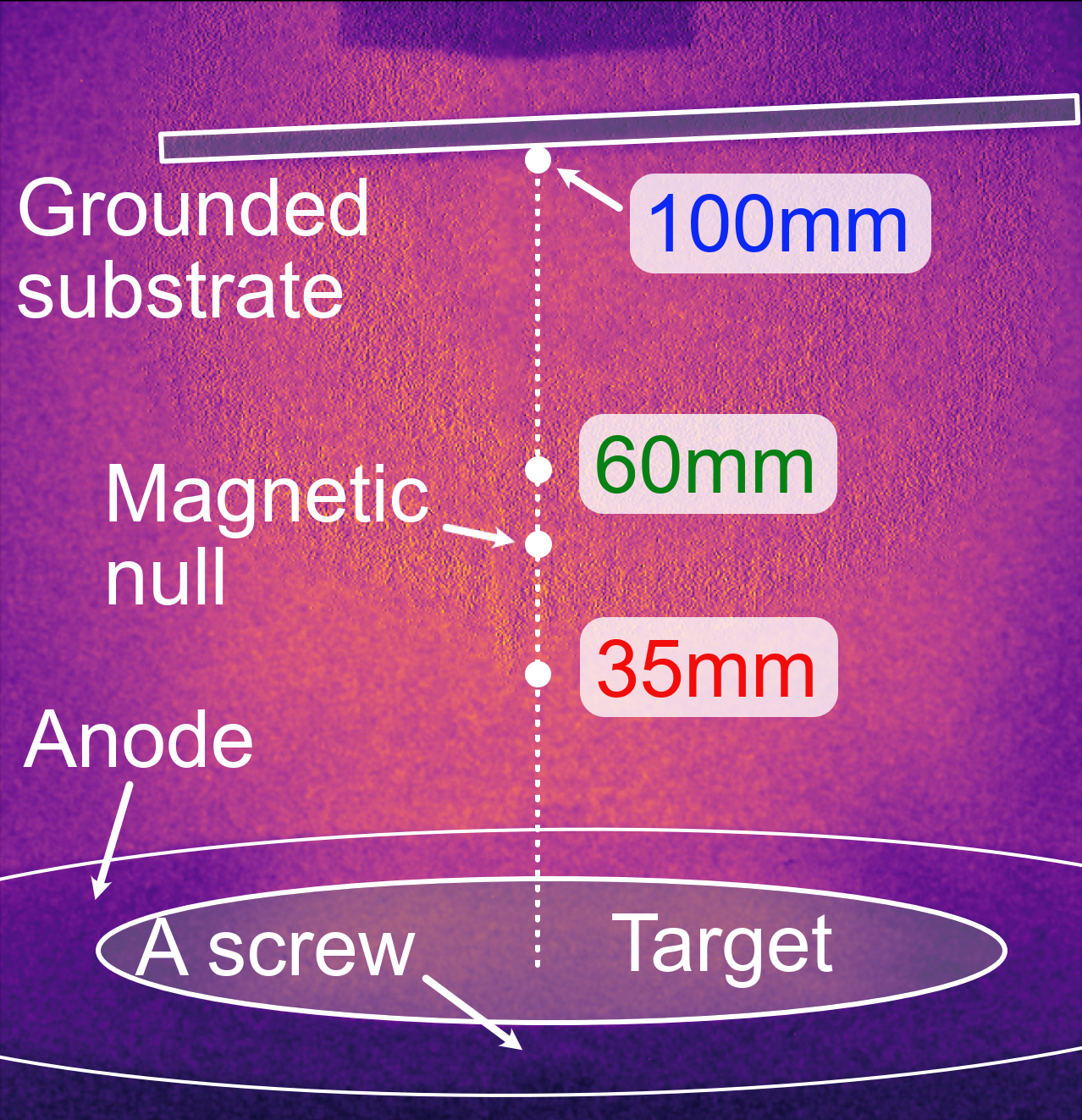}
    \caption{An illustration of the target, anode, and grounded substrate positions on the OES images. The approximate locations of the Langmuir probe tip and the position of the magnetic null are also marked.}
    \label{f:OESimage}
\end{figure}

\Fref{f:OESimage} shows an illustration image from the emICCD camera. The positions of the target, the grounded anode, and the substrate are shown. We want to note the position of a screw holding the anode plate, which may be misleadingly considered the target center. The approximate positions of the probe tip are also marked together with the location of the magnetic null.

\section{Results}

\subsection{Discharge characteristics}

\begin{figure}
    \centering
    \includegraphics{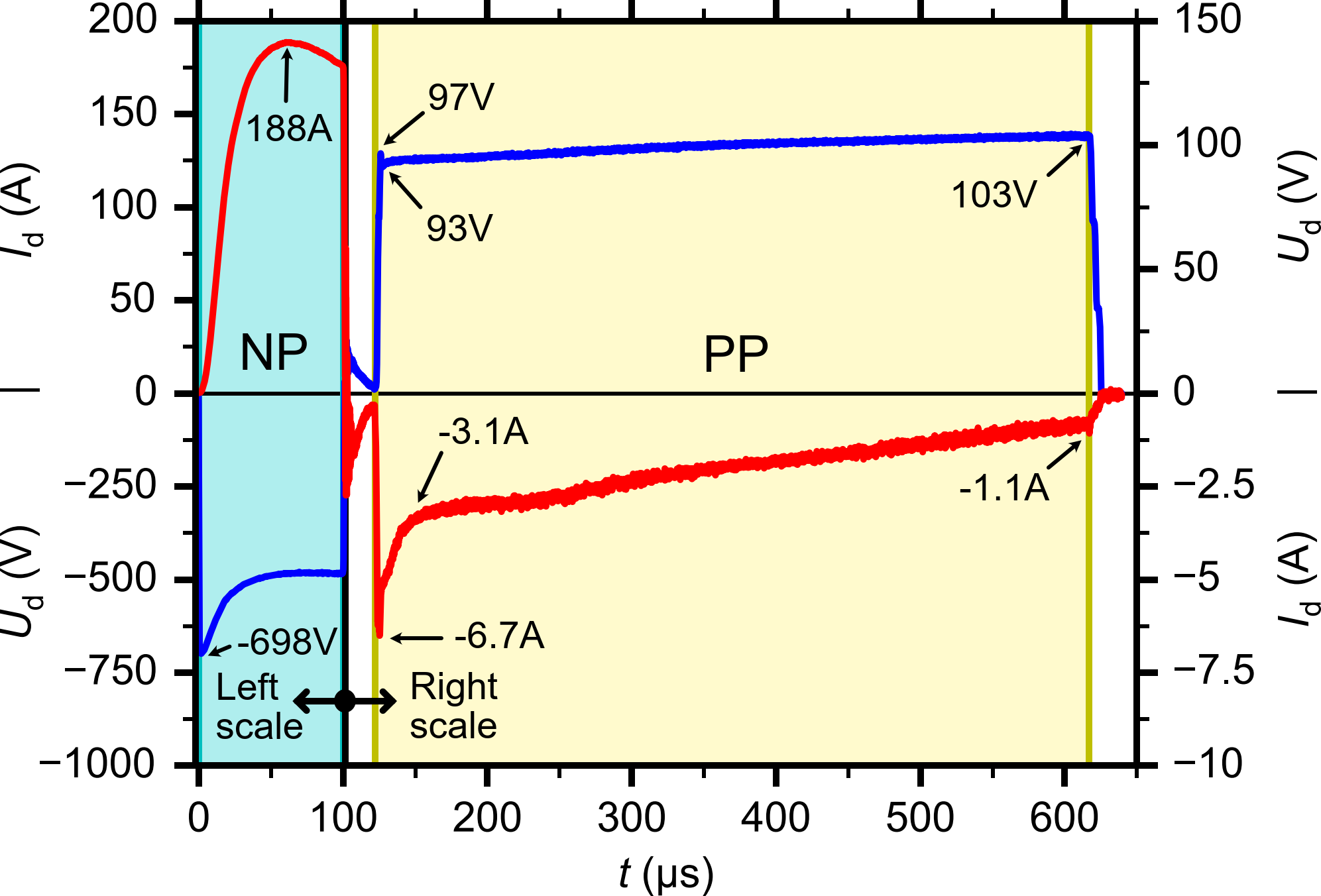}
    \caption{Waveforms of the magnetron voltage, $U_\mathrm{d}$, (blue line), and the discharge current, $I_\mathrm{d}$, (red line) during NP and PP. The left scales hold in front of the black vertical line. Behind this line, the right scales hold.}
    \label{f:UdId}
\end{figure}

\Fref{f:UdId} shows magnetron voltage, $U_\mathrm{d}$, and discharge current, $I_\mathrm{d}$, waveforms during NP and PP. The averaged target power density in NP is $S_\mathrm{da} = 0.96\,\mathrm{kWcm^{-2}}$, and it is calculated from the recorded $U_\mathrm{d}$ and $I_\mathrm{d}$ waveforms by the equation
\begin{equation}
S_\mathrm{da} = \frac{1}{t_-} \int_0^{t_-} U_\mathrm{d}(t) J_\mathrm{d}(t) \, \mathrm{d}t,
\label{e:Sda}
\end{equation}
where $J_\mathrm{d} = I_\mathrm{d} / A_\mathrm{t}$ is the discharge current density and $A_\mathrm{t} = 78.54\,\mathrm{cm^2}$ is the total area of the target. The peak target power density in NP is $S_\mathrm{peak} = 1.16\,\mathrm{kWcm^{-2}}$ at $t = 56\,\mathrm{\mu s}$. Here, it should be noted that the decrease in $U_\mathrm{d}$ during NP is caused mainly by a protective resistor (around $2\,\Omega$) connected up in the output circuit of the pulsing unit (see \fref{f:system}).

After the PP initiation, a steep increase in $U_\mathrm{d}$ leads to a quick rise in $I_\mathrm{d}$ magnitude and the creation of a short $I_\mathrm{d}$ peak. After that, $U_\mathrm{d}$ slightly reduces, and the $I_\mathrm{d}$ magnitude decreases. This behavior is qualitatively similar to those observed in Ref. \cite{Pajdarova2020}, but the values differ. We attribute these differences to a higher erosion of the target as the other discharge parameters are as close as possible to those in Ref. \cite{Pajdarova2020}. This implies that the waveforms of $U_\mathrm{d}$ and $I_\mathrm{d}$ during PP may be susceptible to relatively small variations in the plasma state near the target (the density and spatial distribution of plasma species) before the PP initiation. They are natural consequences of different plasma evolutions during NP as the magnetic field geometry changes with racetrack erosion. After $t \approx 150\,\mathrm{\mu s}$, the decrease of $I_\mathrm{d}$ magnitude slows down, and it continues to decrease up to the end of PP. After the initial overshoot, the value of $U_\mathrm{d}$ increases monotonically during the whole PP due to decreased plasma conductivity and reduced voltage drop across the protective resistor.

\subsection{Local plasma parameters}

\begin{figure}
    \centering
    \includegraphics{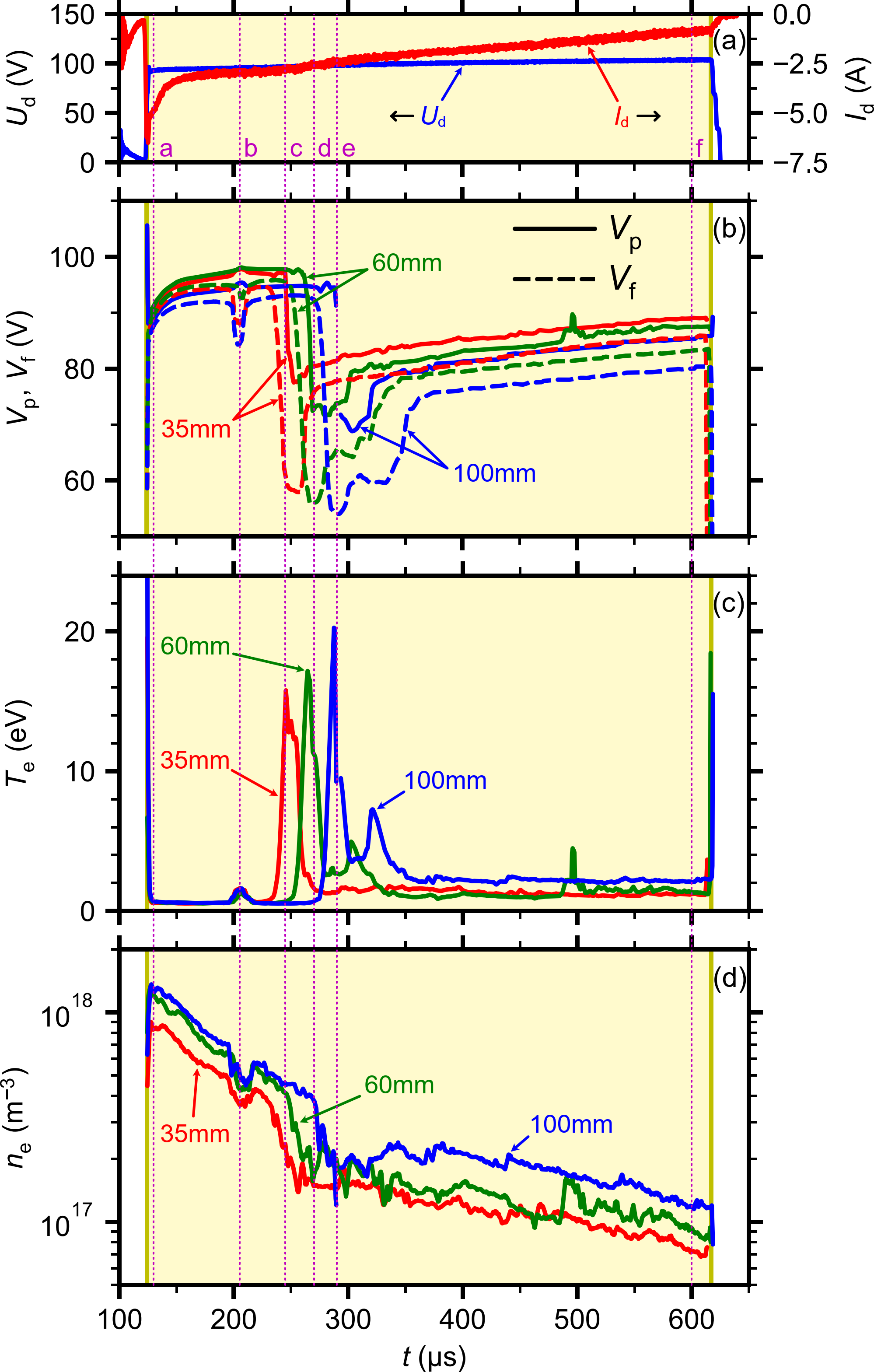}
    \caption{Time evolutions of local plasma parameters during PP at the discharge centerline at the distances $z = 35\,\mathrm{mm}$, $60\,\mathrm{mm}$, and $100\,\mathrm{mm}$ from the target. Panel (a) shows the waveforms of magnetron voltage, $U_\mathrm{d}$, and discharge current, $I_\mathrm{d}$, for comparison purposes. Panel (b) shows time evolutions of the plasma potential, $V_\mathrm{p}$, (full line) and the floating potential, $V_\mathrm{f}$, (broken line). Panels (c) and (d) show time evolutions of the electron temperature, $T_\mathrm{e}$, and the electron density, $n_\mathrm{e}$, respectively. Vertical broken lines a--f mark the OES measurement times of $t_\mathrm{m} = 130\,\mathrm{\mu s}$, $205\,\mathrm{\mu s}$, $245\,\mathrm{\mu s}$, $270\,\mathrm{\mu s}$, $290\,\mathrm{\mu s}$ and $600\,\mathrm{\mu s}$ (see also \fref{f:OES}).}
    \label{f:VpVfTeNe}
\end{figure}

Figure \ref{f:VpVfTeNe} shows time evolution of the plasma potential, $V_\mathrm{p}$, floating potential, $V_\mathrm{f}$, electron temperature, $T_\mathrm{e}$, and electron density, $n_\mathrm{e}$, measured at the discharge centerline at the distances of $35\,\mathrm{mm}$, $60\,\mathrm{mm}$, and $100\,\mathrm{mm}$ from the target. In this subsection, all numerical values of quantities presented in the text are close approximates unless otherwise stated, and the values of both potentials (plasma and floating) are referenced to the ground (see the connection of chamber and all equipment to the ground in \fref{f:system}).

Before the analysis, it should also be noted that even during the pause between the NP end and the PP initiation, which is relatively long in our case ($t_\mathrm{D} = 20\,\mathrm{\mu s}$), the target has a gradually decreasing positive voltage, and a negative current is flowing to the target with a decreasing magnitude. This may influence the initial plasma density \cite{Pajdarova2020}, and it may also significantly change the composition of the ionic particles (Ar$^+$, Ti$^+$, and Ti$^{2+}$) near the target before the PP initiation due to the outflow of these ions.

\subsubsection{Evolution after the PP initiation}
After the PP initiation, a high difference between $V_\mathrm{p}$ and $V_\mathrm{f}$ (up to several tens of volts) persisting for $1\,\mathrm{\mu s}$ is detected at all positions of measurements. This is accompanied by high values of $T_\mathrm{e}$ and temporary decreases in $n_\mathrm{e}$. Similar fast reactions of plasma parameters on the PP initiation were also registered in Refs. \cite{Hippler2020,Law2021,Zanaska2022}. After the initial part, the difference between $V_\mathrm{p}$ and $V_\mathrm{f}$ is relatively small. During this period, $T_\mathrm{e}$ is low, and $n_\mathrm{e}$ almost monotonically decreases. This evolution is similar to the results described in more detail in Ref. \cite{Pajdarova2020} where the plasma parameters in PP with the duration of $50\,\mathrm{\mu s}$ were studied for the same NP parameters. But there is one exception. In this case, the values of $V_\mathrm{p}$ are higher at $z = 60\,\mathrm{mm}$ compared to those at $z = 35\,\mathrm{mm}$. Note, that in Ref. \cite{Pajdarova2020} $V_\mathrm{p}$ was higher at $z = 35\,\mathrm{mm}$ than at $z = 60\,\mathrm{mm}$. This again shows that small changes may relatively extensively influence the plasma parameters in the discharge (we have tried to keep the parameters of NP as close as possible to those in Ref. \cite{Pajdarova2020}). It may indicate that the exact reproduction of phenomena during PP may be problematic.

Around $t_\mathrm{m} = 195\,\mathrm{\mu s}$, a simultaneous decrease in $V_\mathrm{f}$ with a duration of $20\,\mathrm{\mu s}$ at the distances of $z = 35\,\mathrm{mm}$ and $100\,\mathrm{mm}$ (near the target and the substrate, respectively) is detected. At the distance of $z = 60\,\mathrm{mm}$, the decrease in $V_\mathrm{f}$ is also registered, but with a delay of $3\,\mathrm{\mu s}$. These decreases are accompanied by an increase in $T_\mathrm{e}$, a decrease in $n_\mathrm{e}$, and a slight increase in $V_\mathrm{p}$ at the corresponding times. At $t_\mathrm{m} = 215\,\mathrm{\mu s}$, $V_\mathrm{f}$ and $V_\mathrm{p}$ return to their gradual increase from the period before this event. Similarly, $T_\mathrm{e}$ and $n_\mathrm{e}$ also return to the evolution before the $V_\mathrm{f}$ decrease.

A simultaneous decrease in $V_\mathrm{f}$ and rise in $T_\mathrm{e}$ correspond to the classical (equilibrium) theory of plasma floating wall potential \cite{Lieberman2005} considering the elevated values of $V_\mathrm{p}$ (owing to the applied positive potential to the target) leading to the positive $V_\mathrm{f}$. The time evolution of $U_\mathrm{d}$ and $I_\mathrm{d}$ (\fref{f:UdId}) does not exhibit any sign of change in their trends. It means that the origin of this $V_\mathrm{f}$ decrease comes from the internal change in the plasma. Moreover, the flux of the electrons out of the plasma (see the reduction in $n_\mathrm{e}$) does not flow through the target (no visible peak in $I_\mathrm{d}$).

\subsubsection{Drop \& Rise event}
Around $t_\mathrm{m} = 230\,\mathrm{\mu s}$, $V_\mathrm{f}$ near the target ($z = 35\,\mathrm{mm}$) quickly falls. Other researchers also registered similar behavior of potentials \cite{Hippler2020a, Avino2021, Law2021, Mingyue2022, Zanaska2022}, sometimes described as a drop and rise (D\&R) event. The D\&R event moves to higher distances from the target with time. Around $t_\mathrm{m} = 240\,\mathrm{\mu s}$, $V_\mathrm{f}$ drops at $z = 60\,\mathrm{mm}$, and at $t_\mathrm{m} = 265\,\mathrm{\mu s}$, the drop of $V_\mathrm{f}$ is observed at $z = 100\,\mathrm{mm}$. This movement of the D\&R event is not common in the literature because in Refs. \cite{Avino2021, Law2021} the authors see a movement in the opposite direction (from the substrate to the target), and in Ref. \cite{Hippler2020a}, the D\&R event appears at all measured distances simultaneously. Taking the time differences between the points where the $V_\mathrm{f}$ fall stops and the measurement distances from the target, we can obtain that the average speed between the distances $35\,\mathrm{mm}$ and $60\,\mathrm{mm}$ is roughly $1300\,\mathrm{ms^{-1}}$, and between $60$ and $100\,\mathrm{mm}$ the speed is around $2300\,\mathrm{ms^{-1}}$. It indicates that the D\&R event may accelerate during its movement from the target to the substrate.

The drop in $V_\mathrm{f}$ is always followed after a time delay (between $17\,\mathrm{\mu s}$ and $21\,\mathrm{\mu s}$) by a decrease in $V_\mathrm{p}$. The duration of the D\&R events increases with the distance from the target. It is roughly $45\,\mathrm{\mu s}$, $105\,\mathrm{\mu s}$ and $115\,\mathrm{\mu s}$ at distances $z = 35\,\mathrm{mm}$, $60\,\mathrm{mm}$ and $100\,\mathrm{mm}$, respectively, as estimated from the duration of elevated values of the potential differences $V_\mathrm{p} - V_\mathrm{f}$ (not shown). After the end of the D\&R events, both potentials return to their gradual increase, which persists up to the end of PP, but their values are lower than those before the D\&R events by $17\,\mathrm{V}$ on average. This value corresponds well with the energy decrease of the third peak in IEDF measured in the same system \cite{Kozak2020}.

The D\&R events are accompanied by peaks in $T_\mathrm{e}$. At distances $z = 60\,\mathrm{mm}$ and $100\,\mathrm{mm}$, the second peaks in $T_\mathrm{e}$ with lower values are registered approximately $36\,\mathrm{\mu s}$ after the first peaks. Their occurrence corresponds well with the second decrease in $V_\mathrm{f}$ during the D\&R events at these distances. The occurrence of the D\&R events also decreases $n_\mathrm{e}$ values. Contrary to the previously detected decreases around $t_\mathrm{m} = 205\,\mathrm{\mu s}$ where $n_\mathrm{e}$ rises back to almost the values before the drop, after the D\&R events, $n_\mathrm{e}$ stays low. Only at $z = 100\,\mathrm{mm}$, $n_\mathrm{e}$ partially recovers to higher values.

Here, it should be mentioned that an elevation of $T_\mathrm{e}$ and a decrease in $n_\mathrm{e}$ during PP were also proven by laser Thomson scattering \cite{Law2021, Law2023} during BP-HiPIMS of tungsten target when high PP voltages were used ($U_+ \ge 200\,\mathrm{V}$). Contrary to our case, they saw the increase of $T_\mathrm{e}$ to move in the opposite direction (from the substrate to the target) \cite{Law2023}.

\subsubsection{Stabilization of the reverse discharge}
After the D\&R events, $T_\mathrm{e}$ values stay elevated. Around $t_\mathrm{m} = 490\,\mathrm{\mu s}$, $T_\mathrm{e}$ forms a peak in the plasma bulk ($z = 60\,\mathrm{mm}$). This event also manifests itself in the evolution of $V_\mathrm{p}$. From now on, $T_\mathrm{e}$ at $z = 60\,\mathrm{mm}$ is higher than $T_\mathrm{e}$ at $z = 35\,\mathrm{mm}$.

Here, it should be noted that the spatial structure of the potentials changes. Before the D\&R events, the highest $V_\mathrm{p}$ and $V_\mathrm{f}$ values are measured at $z = 60\,\mathrm{mm}$ and the lowest ones at $z = 100\,\mathrm{mm}$. After the D\&R events, the highest values of potentials are detected at $z = 35\,\mathrm{mm}$, and the lowest are again measured at $z = 100\,\mathrm{mm}$.

The D\&R events slow the plasma decay times down approximately four times. The decay times of $n_\mathrm{e}$ were determined from the fits of the equation $n_\mathrm{e}(t) = n_{\mathrm{e}}(t_0) \exp [- (t - t_0) / \tau]$ to the measured $n_\mathrm{e}$ values, where $\tau$ is decay time and $t_0$ is the time from which the fit is performed. The decay times between the PP initiation and the D\&R events (the decreases around $t_\mathrm{m} = 205\,\mathrm{\mu s}$ are not included) are $\tau_1 = 121\,\mathrm{\mu s}$, $112\,\mathrm{\mu s}$ and $113\,\mathrm{\mu s}$ for the distances $z = 35\,\mathrm{mm}$, $60\,\mathrm{mm}$ and $100\,\mathrm{mm}$, respectively, so the average value is $\bar{\tau}_1 = 115\,\mathrm{\mu s}$. After the D\&R events (till the PP end), the decay times increase to $\tau_2 = 466\,\mathrm{\mu s}$, $539\,\mathrm{\mu s}$ and $423\,\mathrm{\mu s}$ for the distances $z = 35\,\mathrm{mm}$, $60\,\mathrm{mm}$ and $100\,\mathrm{mm}$, respectively, giving the average decay time $\bar{\tau}_2 = 476\,\mathrm{\mu s}$. The increase in the decay time is counterintuitive as the magnitude of the discharge current decreases a little bit faster after the D\&R events compared with its evolution closely before the D\&R event occurrence (see \fref{f:UdId}).

\subsection{Light emission of plasma species}

\begin{figure}
    \centering
    \includegraphics{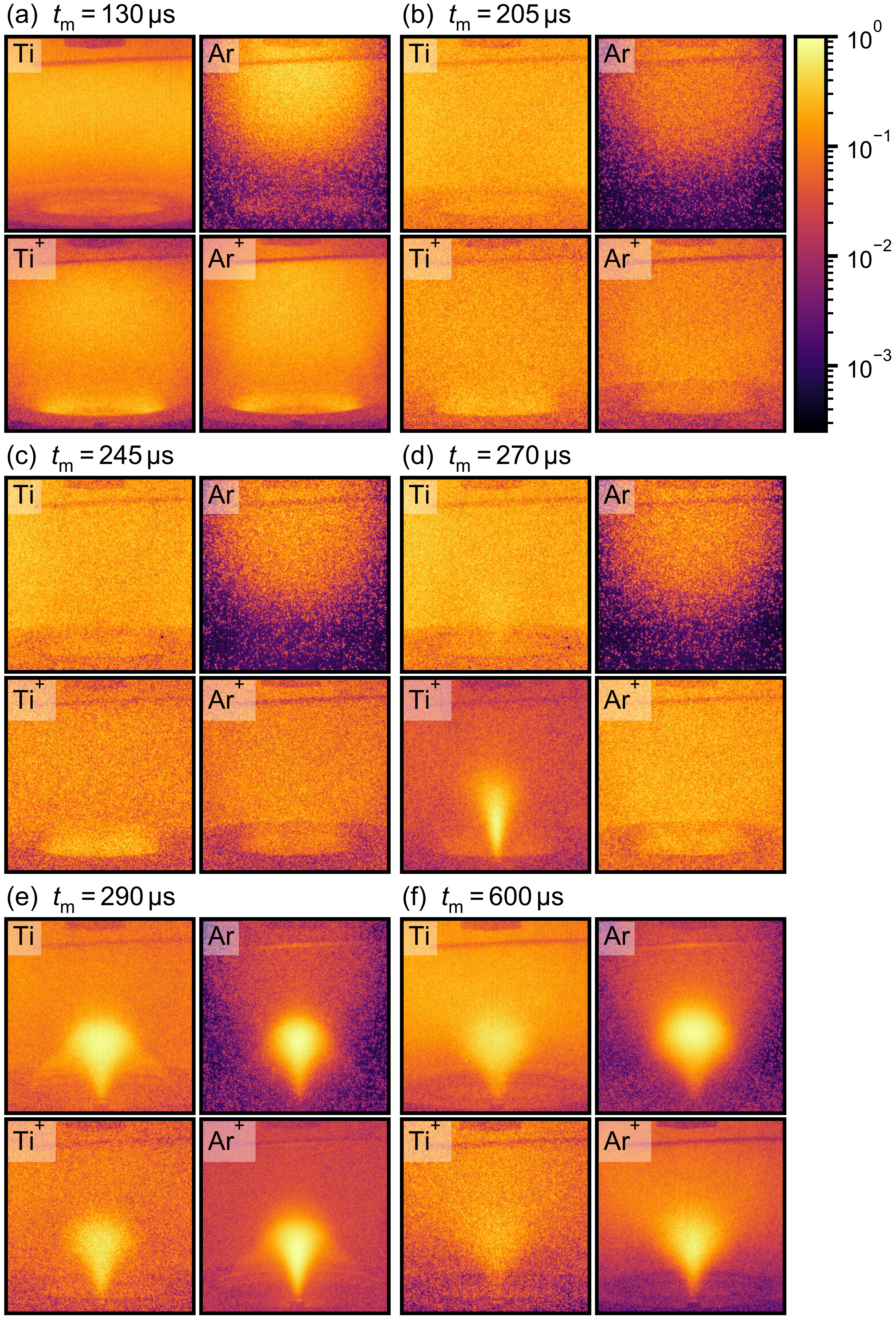}
    \caption{Light emissions from Ti and Ar atoms, and Ti$^+$ and Ar$^+$ ions recorded by emICCD camera for the selected times $t_\mathrm{m} = 130\,\mathrm{\mu s}$, $205\,\mathrm{\mu s}$, $245\,\mathrm{\mu s}$, $270\,\mathrm{\mu s}$, $290\,\mathrm{\mu s}$ and $600\,\mathrm{\mu s}$, which correspond to the vertical broken lines a--f in \fref{f:VpVfTeNe}. The value of 1 corresponds to the most intense emission during PP, and it was calculated independently for each species to emphasize the structure of light patterns. The calibrated intensities of Ar$^+$ ions, Ti atoms, and Ti$^+$ ions are $31.7$, $20$, and $11.2$ times lower than the intensity of Ar atoms.}
    \label{f:OES}
\end{figure}

\begin{figure}
    \centering
    \embedvideo{\includegraphics{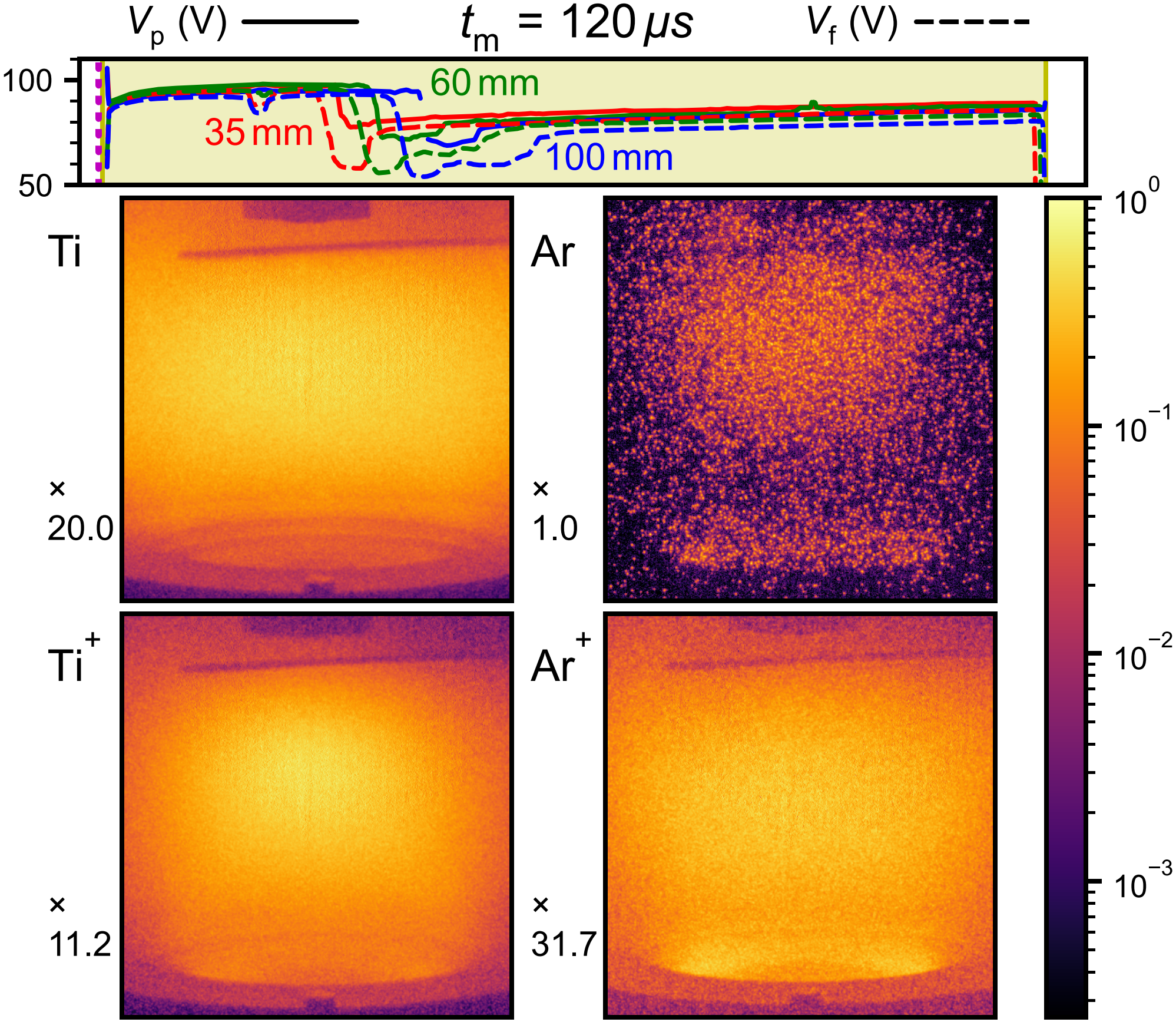}}{video.mp4}
    \caption{Embedded video constructed from all the images captured by the emICCD camera during OES imaging of plasma species (Ti and Ar atoms and singly charged ions) during PP. The multiplication factors for the light intensity are also given for all species to achieve the normalized intensity. Time evolution of the plasma ($V_\mathrm{p}$, full lines) and floating ($V_\mathrm{f}$, broken lines) potentials are given at the probe tip distances from the target $z = 35\,\mathrm{mm}$ (red lines), $60\,\mathrm{mm}$ (green lines), and $100\,\mathrm{mm}$ (blue lines) for reference purposes. The time of measurement is denoted $t_\mathrm{m}$.}
    \label{f:video}
\end{figure}

\Fref{f:OES} shows captured distributions of light intensity originating from the emission lines transmitted by the band-pass filter used for the given specific plasma species (shortly light emission of species) for the selected times $t_\mathrm{m} = 130\,\mathrm{\mu s}$, $205\,\mathrm{\mu s}$, $245\,\mathrm{\mu s}$, $270\,\mathrm{\mu s}$, $290\,\mathrm{\mu s}$, and $600\,\mathrm{\mu s}$, which correspond to the vertical broken lines a--f in \fref{f:VpVfTeNe}. A video constructed from all the images captured by the emICCD camera during PP is embedded in \fref{f:video} and referenced during the following analysis. It should be mentioned that the highest intensity during the PP was registered from Ar atoms, which is consistent with the results in the literature \cite{Kozak2020, Hippler2019, Hippler2020a, Klein2023, Law2023}. The calibrated intensities of other species were lower $31.7$, $20$, and $11.2$ times for Ar$^+$ ions, Ti atoms, and Ti$^+$ ions, respectively, compared to the intensity of Ar atoms.

\subsubsection{Evolution after the PP initiation}
At the beginning of PP (from $t_\mathrm{m} = 120\,\mathrm{\mu s}$ to $150\,\mathrm{\mu s}$), an increased light emission is visible (see \fref{f:OES}a) in the plasma bulk between the magnetic null and the substrate, and close to the target surface as a light ring shifted from the racetrack (RT) position towards the anode. Around $t_\mathrm{m} = 145\,\mathrm{\mu s}$, the intensity of Ti atom-light emission increases in the whole observed volume. It is accompanied by a flash in the intensity of Ti$^+$ ions between $t_\mathrm{m} = 150\,\mathrm{\mu s}$ and $160\,\mathrm{\mu s}$ and a hard-to-see decrease in the intensity of Ar$^+$. During this phase, the distinctly visible rings of ions change to a diffusive light above the target, which lasts up to the D\&R event (see \fref{f:OES}b). The temporary decrease in $V_\mathrm{f}$ around $t_\mathrm{m} = 205\,\mathrm{\mu s}$ does not produce any visible change in the light pattern except that blinking of the species light is evident in the whole plasma volume, which persists up to the D\&R event.

\subsubsection{Drop \& Rise event}
Between $t_\mathrm{m} = 235\,\mathrm{\mu s}$ and $250\,\mathrm{\mu s}$, when $V_\mathrm{f}$ attains its minimum at $z=35\,\mathrm{mm}$, no apparent change in the light emission of species is visible (see \fref{f:OES}c). But, at $t_\mathrm{m} = 255\,\mathrm{\mu s}$, intensity of Ar$^+$ increases in the whole volume. When the D\&R event moves to the distance of $z = 60\,\mathrm{mm}$ at $t_\mathrm{m} = 270\,\mathrm{\mu s}$ (see \fref{f:OES}d), a light pattern resembling a "jet" from the target center is visible in the light emission of Ti$^+$ ions (see \fref{f:OES}d). The beginning of the light pattern formation in the Ti$^+$ ion emission can be seen at $t_\mathrm{m} = 265\,\mathrm{\mu s}$ (see the video in \fref{f:video}). Considering the next video frame ($t_\mathrm{m} = 275\,\mathrm{\mu s}$), it looks like the light pattern expands in the direction from the target center to the magnetic null. According to similar figures acquired in a discharge with a magnetically constricted anode \cite{Chauhan2016}, this is anode light shaped by the presence of magnetic field into the "jet"-like shape. Even though the term anode light is usually used for the full light emitted near the anode, we will use the term anode light pattern (ALP) to describe this shaped light pattern emitted by a given plasma species near the target during PP. Similarly, the inception of ALP is also visible at $t_\mathrm{m} = 270\,\mathrm{\mu s}$ in the light emission of Ti atoms. At $t_\mathrm{m} = 275\,\mathrm{\mu s}$, the ALP of Ti$^+$ ions elongates and becomes more intense. At the same time, the ALP of Ar$^+$ ions appears. Around $t_\mathrm{m} = 280\,\mathrm{\mu s}$, the ALPs of Ti atoms, Ti$^+$ ions, and Ar$^+$ ions are well developed, but the ALP of Ar atoms appears $5\,\mathrm{\mu s}$ later. The appearance of ALPs is also connected with the lowering of the light emission of the corresponding ionic species in the rest of the discharge. Moreover, when the ALP of Ar$^+$ ions develops, the light emission from Ti$^+$ ions in the discharge volume increases again. Between $t_\mathrm{m} = 280$ and $290\,\mathrm{\mu s}$, ALPs start to change their shapes. The body of ALPs becomes wider and flattened near the magnetic null. The light emission from "arcs" connecting the top of the "jets" near the magnetic null and the grounded anode becomes visible. The "arcs" and the "jets" together form ALP resembling an "umbrella". The full development of an ALP for given species (more apparently visible for ions) leads to the diminishing (Ti atoms at $t_\mathrm{m}=280\,\mathrm{\mu s}$) or almost disappearance (Ti$^+$ ions at $t_\mathrm{m}=275\,\mathrm{\mu s}$ and Ar$^+$ ions at $t_\mathrm{m}=280\,\mathrm{\mu s}$) of the diffusive light near the target. Similar light structures were also registered during PP in BP-HiPIMS in Refs. \cite{Kozak2020, Law2023, Klein2023} in the full discharge light or the light emission of Ar atoms.

\subsubsection{Stabilization of the reverse discharge}
Around $t_\mathrm{m} = 360\,\mathrm{\mu s}$, when the potentials at all distances from the target return to their monotonic increase after D\&R events, the emission from the "arcs" between the top of the "jets" and the grounded anode disappears, and ALPs gradually stabilize. Despite the gradual decrease in the discharge current, the Ar and Ar$^+$ ALP intensity stays relatively unchanged. After the PP end, the ALPs of all species disappear during $10\,\mathrm{\mu s}$ and the remaining light from all species gradualy decreases (the fastest decrease is observed in Ar atoms and the remaining Ti$^+$ ions, not shown).

\section{Discussion}

During the analysis of emICCD images, it is important to keep in mind that the registered higher light emission of species provides several pieces of information that must hold simultaneously: (1) the specific species are present in that volume, (2) in the same volume, there are also electrons, and (3) these electrons have enough energy to excite those species. The interpretation of species' light emission lowering is more complicated (it may be a result of ground state density decrease, a decrease of $T_\mathrm{e}$ or $n_\mathrm{e}$, or a combination of causes mentioned above), but it may be unraveled if the light emission decrease of one species is compared with the light emission from the other species. For instance, the decrease in light emission of Ti$^+$ ions and almost constant light emission of Ar$^+$ ions in the same area indicates that the ground state density of Ti$^+$ ions decreases as the electrons populating monitored excited levels of Ar$^+$ ions (upper-level energies of $19.68\,\mathrm{eV}$ and $19.80\,\mathrm{eV}$) will also populate the monitored excited levels of Ti$^+$ with much lower excitation energies (upper-level energies from $3.69\,\mathrm{eV}$ to $4.28\,\mathrm{eV}$ for the lines with dominant emission) when the electron density and temperature does not exhibit substantial changes (they are relatively stable after the stabilization of RD, see \fref{f:VpVfTeNe}).

Let us note that a typical lifetime (considering all transitions leading to the energy level depopulation) of excited radiative states (i.e., excited levels from which spontaneous light emission is allowed) producing high intensities, that are predominantly detected by the emICCD camera, is usually under several hundreds of ns. Typical transition times (connected only with one transition) for the observed emission lines can be calculated from the equation $\tau_{ki} = 1 / A_{ki}$, where $A_{ki}$ is the atomic transition probability from the higher energy level $k$ to the lower energy level $i$. For Ti atom emission lines around $520\,\mathrm{nm}$ these $\tau_{ki}$ are roughly $30$---$270\,\mathrm{ns}$, for Ti$^+$ ion lines around $334\,\mathrm{nm}$ they are $6$---$30\,\mathrm{ns}$, for Ar atom lines around $811\,\mathrm{nm}$ they are $30$---$40\,\mathrm{ns}$, and for Ar$^+$ ion lines around $488\,\mathrm{nm}$ the transition times are $12$---$70\,\mathrm{ns}$ \cite{Kramida2022a}. So, generally speaking, the transition times of emission lines observed in our measurements are sufficiently short to imply that the registered light emission originates almost from the same place where the species were excited, even for ions that are subject to acceleration by the present electric field.

\begin{figure}
    \centering
    \includegraphics{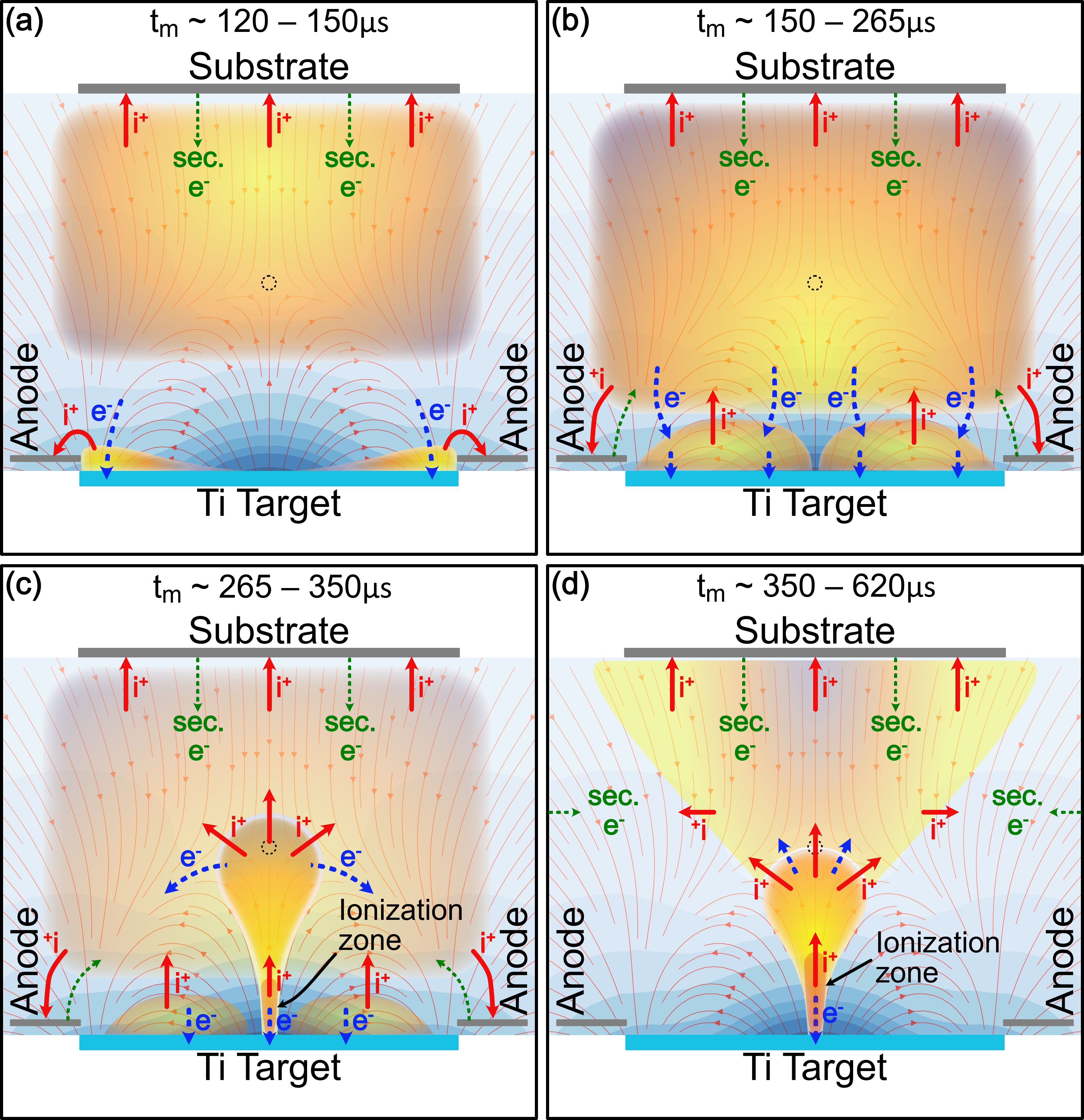}
    \caption{Schematic depiction of the dominant area of plasma radiation and proposed dominant flows of Ti$^+$ and Ar$^+$ ions (red full lines, marked by i$^+$), electrons (blue broken lines, marked by e$^-$), and secondary electrons (green slim broken lines, marked by sec.~e$^-$). The volume of atom ionization is also schematically illustrated in panels (c) and (d).}
    \label{f:fluxes}
\end{figure}

\Fref{f:fluxes} schematically shows the dominant area of the plasma emission during the different phases of PP together with the proposed dominant fluxes of charges. The red full arrows (denoted by i$^+$) represent the fluxes of ions, the blue broken arrows (marked by e$^-$) represent the fluxes of electrons, and the green broken arrows with lower thickness (denoted by sec.~e$^-$) illustrate the places where the electrons are created by the secondary electron emission (SEE) after the ion impacts. In the plasma volume, the flux of one charge is connected with the flux of the charge with the opposite sign owing to the ambipolar diffusion. So, these arrows must be regarded as illustrations of the origin of charge fluxes and the directions in which they are flowing. Also depicted are areas of the discharge where ionization of atoms is possible during the RD ignition and its sustaining.

\subsubsection{Evolution after the PP initiation}
The presence of the light ring near the target edge in the light emission of ions (see \fref{f:fluxes}a) is consistent with the results of ion density measurements after the NP end near the target in HiPIMS discharges \cite{Britun2015a, Pajdarova2022}. A high number of ions and weakly magnetically confined electrons near the target outflow ambipolarly along the discharge centerline, where the magnetic field lines are almost perpendicular to the target surface, to distances beyond the magnetic null already during the pause between the NP end and the PP initiation (as shown in \fref{f:fluxes}a). Only the ions held by magnetically confined electrons stay near the target and create the ring shifted from the position of the deepest RT to the target edges. As the electrons flow to the target, these ions flow to the anode. It was observed for similar pressures in Ref. \cite{Zanaska2022} and predicted by PIC simulations \cite{Han2022}. Ar$^+$ ions located near the substrate (the density of Ti$^{2+}$, which may also induce SEE, is much lower than the density of Ti$^+$ ions \cite{Pajdarova2019}) induce SEE from the substrate surface. Owing to the relatively long delay between NP and PP, the density of ions able to produce secondary electrons (Ar$^+$ and Ti$^{2+}$) is low near the target. Moreover, the density of Ar$^+$ ions near the target was reduced by argon rarefaction (the Ar atom density reduction owing to the momentum transfer from the sputtered metal atoms and their ionization \cite{Gudmundsson2012}) during the high-power NP. Thus, the production of secondary electrons near the target is low.

The video (see \fref{f:video}) shows that the Ti$^+$ ion ring near the target edge increases its intensity after the PP initiation, which is connected with the increased density of electrons in this location. The ion ring is the most probable location where the electron current flows to the target (see \fref{f:fluxes}a), as the missing emission from all species near the target center supports this scenario. The electrons flowing to the target also drag the ambipolar flow of ions with them, and the plasma region starts to extend back to the target (\fref{f:fluxes}b). The ion ring near the target edge starts to disappear, which is also observed in the PIC simulation \cite{Han2022}. This is also accompanied by a widening of the volume of positive plasma potential near the target. Despite the positive plasma potential close to the target, Ti$^+$ ions are held in the magnetic trap by the magnetically confined electrons, but the ions are repelled to higher distances from the target, and they may also, to some extent, escape in the direction toward the substrate (more visible for Ar$^+$ ions, see \fref{f:OES}b). The widening of the positive $V_\mathrm{p}$ volume also increases the volume from which electrons are drawn to the target vicinity. This also leads to the enlargement of the target area, where its surface absorbs electrons (see \fref{f:fluxes}b). It should be noted that these changes in light emission coincide with the change in the decrease rate of $I_\mathrm{d}$, but these changes are not visible in the Langmuir probe measurements.

As the temporary decrease in $V_\mathrm{f}$ around $t_\mathrm{m} = 205\,\mathrm{\mu s}$ does not produce any changes in the registered light emission, we have to state here that the origin of this decrease is unknown.

\subsubsection{Drop \& Rise event}
As the target potential is still close to $U_+ = 100\,\mathrm{V}$ (see \fref{f:UdId}), but the plasma potential at $z = 35\,\mathrm{mm}$ drops down to $59\,\mathrm{V}$, the DL structure must be formed at closer distances to the target. The PIC simulations \cite{Avino2021, Han2022} show that the volume of more positive $V_\mathrm{p}$ concentrates around the discharge centerline near the target in the form of a "hat" with the crown oriented in the direction toward the substrate and with the brim settling on the target. This volume of the positive $V_\mathrm{p}$ extends roughly $10\,\mathrm{mm}$ from the target surface \cite{Avino2021}, and beyond this distance, $V_\mathrm{p}$ quickly decreases.

ALP is formed around the discharge centerline (see figures \ref{f:OES}d-e and \ref{f:fluxes}c) firstly in the light emission of Ti$^+$ ions, then it is also found in the light emission of other species. As mentioned above, similar light patterns (in the full discharge light) were found in the discharges with the magnetically constricted anode \cite{Chauhan2016}, where a positive voltage is applied on the magnetron (the anode) and the chamber walls are grounded (the cathode). The authors explain this light pattern as anode light (sometimes also called a fireball) deformed by the presence of the magnetron's magnetic field that constricts the anode area. As the area at the anode, where the electron current is drawn, is substantially smaller than the area of the cathode, the electron sheath \cite{Scheiner2015, Sun2022} is formed above this anode spot. The electron sheath then allows the electron current to reach values greater than the thermal electron current (given by the electron saturation current in \eref{e:Ie}).

It should be noted here that the magnetic field geometry forms a magnetic mirror for the electrons flowing to the target through the magnetic funnel. It is complicated to correctly describe (in all detail) its effect in BP-HiPIMS, as the magnetic mirrors are usually studied without the presence of electric field and particle collisions. We can speculate that the electric field accelerates the electrons in the direction toward the target, so the velocity component, which is parallel with the magnetic field lines (they are here almost perpendicular to the target), increases. This pushes the electron velocity inside the loss cone \cite{Dinklage2005a} (defined for a given charged particle as the maximum angle between its velocity and the magnetic field line for which the particle can go through the magnetic mirror), and the electrons more easily reach the target through the magnetic funnel. On the other hand, the collisions of electrons with atoms and ions scatter the electron velocity orientation, which results in a higher number of electron reflections. Generally speaking, some electrons go through the magnetic mirror and form the electron current through the center of the target. Others bounce back, increasing the local electron density in the magnetic funnel. This increase may be so high that the plasma quasineutrality may be violated \cite{Avino2021}. It should be noted that a small dimple in the central part of the positive potential "hat" is visible in simulation results \cite{Han2022}, which supports our explanation. The presence of the magnetic mirror can also explain why the electron current flows through the target edge after the PP initiation (see \fref{f:fluxes}a) as the flow through the target center is suppressed by the magnetic mirror. For the full description, the theoretical treatment or detailed simulation of the electron sheath in the magnetic mirror configuration would be needed, but this is out of the scope of this paper.

Although the electron sheath can collapse into a DL structure itself as numerically proven \cite{Conde2006} without the magnetic field presence, the magnetic mirror configuration may significantly accelerate the creation of the DL structure. The formation of the DL structure originates from the plasma's tendency to shield the target's positive potential. Usually, the electrons are the species that start the shielding owing to their very low mass, but in the case of the magnetic field present, their mobility across magnetic field lines is limited much more than for ions, and the magnetic mirror effect limits their mobility through the magnetic funnel. Thus, the electrons surround the volume where the magnetic trap is located. Ions may, to some extent, freely outflow in the direction of the substrate, which strengthens the electron density behind the magnetic trap, and the plasma potential is lowered here. It leads to the formation of the DL structure.

Since the electrons gain energy when they cross the DL boundary from its L-side to the H-side and the magnetic mirror increases their density close to the target central position, the probability of atom ionization in the magnetic funnel increases. Because the ionization energy of Ti atoms is much lower than that of Ar, the ALP is first seen in the light emission of Ti$^+$ ions. Let us note that the density of Ti atoms is relatively high near the target. The ionization of Ti atoms increases the density of electrons in the magnetic funnel even more. The increased density of electrons makes it possible to start the ionization of Ar atoms with much higher ionization energy than Ti atoms. Moreover, the density of Ar atoms had enough time to recover from its decrease caused by their rarefaction during NP as the times for the density replenishing near the target are roughly $100$ -- $150\,\mathrm{\mu s}$ \cite{Vitelaru2012, Kozak2018}. Thus, the ALP of Ar atoms and ions is visible later. The growth of the ALP from the target to the magnetic null is caused partly by the ambipolar flow of the newly created ions, partly by the reflection of the new electrons by the magnetic mirror, and partly by the inflow of new electrons from the plasma bulk. Here it should be noted that narrower and sharper ALPs registered in the case of ionic species (see \fref{f:OES}d--f) cast doubt on the proposed radial outflow of ions in the electron sheath of the discharge with the magnetically constricted anode \cite{Chauhan2016}. All these phenomena lead to an increased density of electrons from the target surface up to the position of the magnetic null. This was also observed in PIC simulations \cite{Avino2021}. These electrons excite species located around the discharge axis, which then emit light. Moreover, these electrons follow the magnetic field lines and create "arcs" in the "umbrella" shape light pattern. Some electrons in the "arcs" also originate from the secondary emission after the newly created Ar$^+$ ions hit the anode. The inflow of new ions to the plasma gradually increases the plasma potential from inside the magnetic funnel up to the substrate. That explains the rise of $V_\mathrm{p}$ after its decrease during the initial part of the D\&R event. From this time, the RD can be regarded as fully developed because the newly created Ar$^+$ ions can produce secondary electrons from the grounded surfaces.

\subsubsection{Stabilization of the reverse discharge}
Now, the RD is burning between the substrate, which is the source of secondary electrons emitted mainly after Ar$^+$ ion impacts, and the target center, where the electrons are collected (see \fref{f:fluxes}(d)). Afterward, the density of Ti$^+$ ions gradually decreases (see the stable intensity of Ar$^+$ in \fref{f:OES}e--f and the video) as they outflow to the grounded substrate and walls. The ionization of Ti atoms in the ionization zone of RD (see \fref{f:fluxes}d) cannot replenish the decrease of Ti$^+$ density as the target is not sputtered. Ar$^+$ ions, created by the ionization of Ar atoms in the ionization zone, start to play a dominant role in RD. This also means that the substrate is dominantly bombarded by Ar$^+$ ions with elevated energies during RD. Despite the gradual decrease in the discharge current, Ar and Ar$^+$ ALP intensities stay relatively unchanged. It shows that RD can generate a sufficient density of Ar$^+$ ions to sustain RD by the secondary electrons emitted mainly from the substrate after the Ar$^+$ ions impact it.

\subsection{Conditions for the creation of a double-layer structure and reverse discharge}
Let us summarize which conditions would accelerate the formation of the DL structure and the ignition of RD in BP-HiPIMS discharges during PP. One condition is an effective magnetic mirror in front of the target, but this condition is fulfilled in almost all magnetron discharges. Magnetrons with a balanced magnetic field would be more effective in the electron accumulation behind the magnetic trap. In those cases (if the construction of the magnetron is classical with the protruding anode ring around the target), electrons behind the magnetic trap cannot easily flow to the target edge as they are repelled by the grounded anode, which is close to or even in the path of the magnetic field lines leading electrons to the target edge (see \fref{f:fluxes}a). In these cases, electrons must diffuse across the magnetic field lines or go directly through the magnetic mirror in the magnetic funnel. The second condition is to have ionic species capable of effective SEE from the grounded surfaces near the target to build up the electron density by the SEE in front of the magnetic trap. This means that PP must be initialized almost immediately after the NP end as it prevents the outflow of Ar$^+$ ions and doubly-ionized metal atoms to higher distances from the target or even onto the grounded surfaces near the target. The shorter NP duration can also increase the probability of the DL structure formation as a lower Ar density reduction (weaker rarefaction) during NP allows Ar$^+$ ions to have a higher density close to the target at the PP initiation. Moreover, if the power during the NP is high, a relatively high density of doubly-ionized metal ions may be accumulated in front of the target. Smaller magnetron sizes also boost the DL formation since the diffusion lengths of the ions inducing SEE from the grounded surfaces near the target are shorter. This is why the DL formation during relatively short times is often observed in small-size balanced magnetrons driven by a relatively short NP with a minimal delay between the NP end and the PP initiation, e.g., as seen in Refs. \cite{Velicu2019, Keraudy2019, Tiron2020}.

The creation of DL and the ignition of RD during PP may be beneficial for film formation if the ignition of RD is done almost immediately after the PP initiation. Sputtered metallic atoms during NP would be ionized in the RD ionization zone, and they may gain energy in the DL structure when they cross from the H-side with a higher potential to the L-side with a lower potential. Unfortunately, as the metallic atoms are not replenished by sputtering from the target, RD relatively quickly starts to ionize Ar atoms, and the deposited film is bombarded by energetic Ar$^+$ ions that may induce defects in the growing film.

\section{Conclusions}

The time-resolved Langmuir probe diagnostics and optical emission spectroscopy imaging for different plasma species (Ti and Ar atoms as well as singly ionized Ti and Ar ions) have been carried out during the long positive voltage pulse (duration of $500\,\mathrm{\mu s}$) in bipolar HiPIMS discharge (with the positive voltage of $100\,\mathrm{V}$). Comparing our results with those in the literature makes it possible to identify the main phenomena leading to the formation of the double-layer structure and the reverse discharge ignition. Our findings may be summarized in these points:

\begin{itemize}
    \item After the positive voltage pulse initiation, the high difference between the values of the plasma potential and the floating potential is registered, which persists only around $1\,\mathrm{\mu s}$, and it is accompanied by a high electron temperature (almost up to $50\,\mathrm{eV}$). After that, both potentials monotonically increase (except for a short time when the temporal decrease of the floating potential is registered). The light emission from the plasma species does not show any unexpected phenomena.
    \item Roughly at the second quarter of the positive voltage pulse, a decrease in the floating potential (by up to $40\,\mathrm{V}$) followed after a few $\mathrm{\mu s}$ by a decrease in the plasma potential (by up to $26\,\mathrm{V}$) in a large volume of the discharge plasma is observed. This is accompanied by large peaks in the electron temperature (up to $20\,\mathrm{eV}$) and an elongation of the electron density decay times (from $115$ to $476\,\mathrm{\mu s}$). The changes in the plasma parameters are followed by the presence of anode light patterns located between the target center and the magnetic null of the magnetron at the discharge centerline. The reverse discharge is ignited.
    \item After between $45\,\mathrm{\mu s}$ and $115\,\mathrm{\mu s}$ depending on the distance from the target, the plasma, and floating potentials return to their increases from the times before their sudden decrease, but the values of the potentials are lowered compared to those before the decrease (by $17\,\mathrm{V}$ on average). The electron temperatures rise (up to $2\,\mathrm{eV}$ in comparison to values up to $0.2\,\mathrm{eV}$ from the initial part of the positive pulse.). The anode light patterns stabilize and persist until the positive voltage pulse ends. The intensity of the light patterns of Ti ions and atoms gradually decreases as their densities decrease and are not replenished by the sputtering from the target. After the positive voltage pulse ends, these light patterns vanish entirely.
    \item The secondary electron emission induced dominantly by Ar$^+$ ions striking the grounded surfaces and the mirror effect of the magnetron magnetic field were identified as probable causes of the charge double-layer structure creation and the maintenance of the reverse discharge. The influx of the secondary electrons to the target vicinity induces the plasma potential decrease behind the magnetic trap, which results in a quicker formation of the double-layer structure in front of the target. The increased electron density near the target at the discharge centerline, owing to the magnetic mirror effect and the energy that electrons gain during their transition through the double-layer structure, allows starting the ionization of atoms and the ignition of the reverse discharge.
    \item Since the reverse discharge is burning mainly due to the creation of Ar$^+$ ions that can supply the reverse discharge with a sufficient number of secondary electrons emitted from the grounded substrate after the Ar$^+$ ion impact, the growing film on the substrate is bombarded during the reverse discharge mainly by these ions with elevated energies. It may lead to the creation of defects or even resputtering of the growing film. Thus, the reverse discharge should be avoided when high-quality, densified films are requested. On the other hand, when porous films should be deposited, reverse discharge may be beneficial.
\end{itemize}

\ack
This work was supported by the project QM4ST with No. CZ.02.01.01/00/22\_008/0004572, co-funded by the ERDF as part of the Ministry of Education, Youth and Sport.

\section*{References}
\bibliographystyle{unsrt}
\bibliography{references}

\end{document}